\documentclass[lettersize,journal]{IEEEtran}
\usepackage[noadjust]{cite}
\usepackage{amsmath,amssymb}
\usepackage{graphicx}
\usepackage{braket}
\usepackage{epsfig}
\usepackage{epstopdf}
\usepackage{mdwmath}
\usepackage{xcolor}
\usepackage{varwidth}
\usepackage{framed}
\usepackage{float}
\usepackage{placeins}
\usepackage{afterpage}
\usepackage{ragged2e}
\usepackage{amssymb}
\usepackage{pifont}

\usepackage{blindtext}
\usepackage{enumitem}
\usepackage{xcolor}
\usepackage{eqparbox}
\usepackage{fixltx2e}
\usepackage{multirow}
\usepackage{enumitem,lipsum}
\usepackage[utf8]{inputenc}
\usepackage[english]{babel}
\newcolumntype{P}[1]{>{\centering\arraybackslash}p{#1}}
\usepackage{graphicx}
\usepackage{amsmath, amssymb}
\usepackage{graphicx}
\usepackage{epsfig}
\usepackage{epstopdf}
\usepackage{array}
\usepackage{graphicx}
\usepackage{subcaption}
\usepackage{hyphenat}
\usepackage[english]{babel}
\usepackage[utf8]{inputenc}
\usepackage{algorithm}
\usepackage{physics}
\usepackage[noend]{algpseudocode}
\usepackage{cuted}
\usepackage{enumitem}
\usepackage{todonotes}
\usepackage[
  group-separator = {\,},
  group-minimum-digits = 4
]{siunitx}

\usepackage{booktabs} 
\usepackage{graphicx}
\usepackage{caption}
\usepackage{bbding}
\usepackage{pifont}
\usepackage{wasysym}
\usepackage{amssymb}

\usepackage{framed,url,caption,graphicx}
\usepackage{amsmath,xcolor,soul,algpseudocode,algorithm,xspace,tabularx,multirow}
\usepackage{amssymb, cleveref}

\PassOptionsToPackage{hyphens}{url}

\newcommand{\squishenum}{   \begin{enumerate}{}    { \setlength{\itemsep}{0pt}      \setlength{\parsep}{0pt}      \setlength{\topsep}{3pt}       \setlength{\partopsep}{0pt}      \setlength{\leftmargin}{1.5em} \setlength{\labelwidth}{1em}      \setlength{\labelsep}{0.5em} } }
\newcommand{\squishlist}{   \begin{list}{$\bullet$}    { \setlength{\itemsep}{0pt}      \setlength{\parsep}{3pt}      \setlength{\topsep}{3pt}       \setlength{\partopsep}{0pt}      \setlength{\leftmargin}{1.5em} \setlength{\labelwidth}{1em}      \setlength{\labelsep}{0.5em} } }
\newcommand{\squishlisttwo}{   \begin{list}{$\bullet$}    { \setlength{\itemsep}{0pt}    \setlength{\parsep}{0pt}      \setlength{\topsep}{0pt}     \setlength{\partopsep}{0pt}      \setlength{\leftmargin}{2em} \setlength{\labelwidth}{1.5em}      \setlength{\labelsep}{0.5em} } }
\newcommand{\squishend}{    \end{list}  }
\newcommand{\squishenumend}{	\end{enumerate}	}
\usepackage[utf8]{inputenc}
\usepackage[english]{babel}
\usepackage{fancyhdr}

\usepackage{parselines} 
\usepackage{blindtext}

\usepackage{xcolor}
\usepackage{mathtools}

\newcommand{\eat}[1]{}

\title{Scalable and Efficient Intra- and Inter-node Interconnection Networks for Post-Exascale Supercomputers and Data centers
\thanks{This work is supported by the Spanish Ministry of Science and Universities MCIN/AEI/10.13039/501100011033, the European Union (NextGenerationEU/PRTR) under project TED2021-130233B-C31, the Junta de Comunidades de Castilla-La Mancha and FEDER funds under the project SBPLY/21/180501/000248, the PERTE-Chip grants (UCLM Chair, TSI-069100-2023-0014) funded by the Spanish Ministry of Digital Transformation and Public Service, and the Universidad de Castilla-La Mancha under project 2023-GRIN-34056.}
}

\author{
Joaquin Tarraga-Moreno\textsuperscript{*},
Daniel Barley\textsuperscript{\dag},
Francisco J. Andújar Muñoz\textsuperscript{\ddag},
Jesus Escudero-Sahuquillo\textsuperscript{*},
Holger Fröning\textsuperscript{\dag},
Pedro Javier Garcia\textsuperscript{*},
Francisco J. Quiles\textsuperscript{*},
José Duato\textsuperscript{\S}
\thanks{\textsuperscript{*}Department of Computing Systems, Universidad de Castilla-La Mancha, Spain.}
\thanks{\textsuperscript{\dag}Institute of Computer Engineering (ZITI), Heidelberg University.}
\thanks{\textsuperscript{\ddag}Department of Computer Science, Universidad de Valladolid}
\thanks{\textsuperscript{\S}Openchip \& Software Technologies}
\thanks{E-mail(s): antonioj.tarraga@uclm.es}
}
\begin{document}
\maketitle

\begin{abstract}
The rapid growth of data-intensive applications such as generative AI, scientific simulations, and large-scale analytics is driving modern supercomputers and data centers toward increasingly heterogeneous and tightly integrated architectures. These systems combine powerful CPUs and accelerators with emerging high-bandwidth memory and storage technologies to reduce data movement and improve computational efficiency. However, as the number of accelerators per node increases, communication bottlenecks emerge both within and between nodes, particularly when network resources are shared among heterogeneous components.

This paper analyzes the scalability challenges of such architectures and evaluates communication strategies that address the growing bandwidth and latency demands of accelerator-rich environments. We focus on the NVIDIA DGX GH200 system, which employs Grace Hopper superchips interconnected through a two-level NVLink-based slimmed fat-tree topology. Using traffic simulations under random all-to-all workloads, we evaluate multiple system configurations and compare them against reference InfiniBand (IB-NDR400) networks. Results show that the DGX GH200 interconnect achieves up to 450 Tbps of throughput—substantially outperforming traditional RLFT networks. These findings highlight the effectiveness of dedicated accelerator networks in overcoming communication bottlenecks in next-generation high-performance computing systems.

\end{abstract}

\section{Introduction}

Modern supercomputers and data centers are composed of thousands of interconnected server nodes that continuously expand their computing, memory, and storage capabilities to satisfy the growing demands of data-intensive applications such as scientific simulations, generative AI models, and large-scale social networks. While the demand for computational power is largely being met through increasingly powerful processors and specialized accelerators (e.g., GPUs and TPUs), emerging memory and storage technologies—such as 3D-stacked high-bandwidth memory (HBM), Non-Volatile Memory Express (NVMe), and Storage Class Memory (SCM)—are revolutionizing the architecture of these systems. By providing higher bandwidth and capacity, these technologies enable unprecedented volumes of data to be stored closer to the processing units, significantly reducing data movement and inter-node communication latency. As a result, more computation can be performed locally and efficiently within each server node.

To keep pace with these developments, modern supercomputers are evolving both by scaling up—increasing the computational density within individual nodes—and by scaling out—expanding the number of interconnected nodes. A key challenge in this evolution is the growing need for faster intra- and inter-node communication, particularly among accelerators that frequently exchange large volumes of data. Companies such as NVIDIA are addressing this challenge by developing architectures that feature high-bandwidth, low-latency communication fabrics dedicated to tightly coupled accelerators, improving performance for workloads that rely on rapid local data exchange.

However, as systems continue to scale, relying solely on a shared network infrastructure for all components can create communication bottlenecks. To alleviate this, modern cluster designs are increasingly adopting hierarchical or multi-tiered networking architectures, where dedicated high-speed networks interconnect specific components—such as accelerators—independently from the general-purpose system network. While this approach improves performance isolation and reduces contention, it also introduces significant complexity and cost, as it requires additional links, switches, and network management resources.

At the node level, heterogeneous architectures are also becoming the norm, with systems integrating multiple types of accelerators to optimize different workloads. For instance, the JUPITER supercomputer~\cite{JUPITER} (\#4 of Top500 list) incorporates four GPUs per compute node to achieve higher on-node parallelism and throughput. However, this scale-up approach presents new challenges: as the number of accelerators per node increases, so does the volume of intra-node and inter-node traffic. If the network interface controller (NIC) remains a single shared communication channel, it may become a performance bottleneck, limiting overall scalability and efficiency.

This work addresses one of the main challenges in supercomputers and data centers: \emph{how do accelerators (such as GPUs, TPUs, and others) communicate with each other?} We analyze several real-world systems, including the NVIDIA DGX GH200, to study their communication architectures and propose new intra- and inter-node network topologies aimed at improving overall system performance.

\section{Background}
\label{sec:background}

This section provides background details on a real system architecture proposed by a company, such as NVIDIA. Moreover, we analyze slimmed fat-tree topologies and their routing.

\subsection{DGX GH200 system}
\label{sec:background:dgx}

The DGX GH200 system~\cite{dgx_gh200} proposed an AI-focused system using the NVIDIA Grace Hopper GH200 superchips~\cite{gracehopper} architecture. NVIDIA GH200 Grace Hopper Superchip proposed an unified CPU–GPU architecture that merges an Arm-based NVIDIA Grace CPU with an NVIDIA Hopper GPU into a single package. The two components are connected via NVLink-C2C~\cite{nvlinkC2C}, a high-bandwidth, low-latency, and memory-coherent interface delivering \num{900} GB/s of bidirectional bandwidth—seven times faster than PCIe Gen5.

This superchip architecture combines up to \num{72} Arm Neoverse V2 CPU cores, \num{96} GB of HBM3 GPU memory (\num{4} TB/s bandwidth), and up to \num{480} GB of LPDDR5X system memory (\num{500} GB/s bandwidth), forming a heterogeneous computing unit optimized for AI, HPC, and data analytics workloads. The unified memory access model allows both CPU and GPU to share and access each other’s memory coherently, dramatically reducing data movement overhead.

To scale beyond a single GH200 superchip, NVIDIA employs the NVLink Switch System~\cite{nvlink}, powered by fourth-generation NVLink and third-generation NVSwitch ASICs. Each NVLink Switch provides \num{25.6} Tb/s of full-duplex bandwidth, exposing \num{128} fourth-generation NVLink ports. The NVLink Switch System forms a two-level, non-blocking slimmed fat-tree topology connecting up to \num{256} GH200 superchips. At the first level, groups of eight Grace Hopper modules are interconnected within a chassis by three NVLink Switch trays, providing \num{3.6} TB/s of intra-chassis bisection bandwidth. At the second level, \num{36} NVLink switches interconnect \num{32} of these chassis to form the complete DGX GH200 fabric, offering \num{115.2} TB/s of bisection bandwidth—over nine times higher than the bandwidth of an NDR400 InfiniBand fabric. This topology ensures that every GPU in the system can access the memory of any other GPU or CPU with minimal latency, creating a single, unified high-speed memory space across all \num{256} superchips.

\subsection{Slimmed fat-tree}
\label{sec:background:fattree}

Fat-tree topologies~\cite{6312192} are a class of hierarchical interconnection networks widely used in high-performance computing (HPC) and large-scale data centers due to their scalability, high throughput, and path redundancy. The fat-tree was designed as a universal network providing the same bandwidth at all levels of the hierarchy by extending links toward the root of the tree to avoid bottlenecks and achieve full bisection bandwidth across all communication paths.

Over the years, several generalizations of the original fat-tree have been proposed, including k-ary n-trees, generalized fat-trees (GFTs), and extended generalized fat-trees (XGFTs). These models provide a flexible mathematical framework to represent multilevel switch networks with configurable degrees of connectivity and oversubscription. Fat-trees have been implemented in many supercomputers and commercial interconnects such as InfiniBand due to their deterministic routing, non-blocking behavior, and hardware efficiency.

Because traditional fat-trees aim for full bisection bandwidth, they often require a large number of switches and cables at the upper levels, which significantly increases cost and power consumption. This has motivated the development of slimmed fat-trees, a cost-reduced variation that maintains acceptable performance under realistic HPC workloads that rarely saturate the network’s full capacity. Studies~\cite{4663766,1526015} have shown that many HPC applications underutilize this bandwidth, motivating the development of slimmed fat-trees — cost-effective variants in which upper-level links are oversubscribed without significantly degrading performance.

While traditional fat-trees provide full bisection bandwidth by maintaining an equal number of upward and downward links at each switch level, such configurations are often overprovisioned for the communication demands of real HPC workloads. To reduce cost, power consumption, and cabling complexity, researchers have proposed slimmed fat-trees, in which the upper levels of the topology are oversubscribed—that is, they contain fewer links or switch ports connecting to higher levels than would be required for full bisection bandwidth. Slimmed fat-trees have been adopted in several large-scale supercomputers, including Stampede at the Texas Advanced Computing Center~\cite{Stampede3}, demonstrating the viability of such designs in production HPC environments.

In such configurations, traditional static routing schemes such as destination-mod-k (D-mod-k) and source-mod-k (S-mod-k), which achieve perfect load balance in full-bandwidth fat-trees, tend to produce load imbalance in slimmed fat-trees. This imbalance results in uneven utilization of network links, creating potential bottlenecks and reducing overall communication efficiency. To address this limitation, proposed Round-Robin Routing (RRR)~\cite{YUAN20142423}, a deterministic, traffic-oblivious routing algorithm that aims to distribute source–destination (SD) pairs across all links in extended generalized fat-trees (XGFTs). RRR achieves near-perfect load balancing for two- and three-level XGFTs, covering both traditional and slimmed fat-tree topologies.

\section{Modeled Network}
\label{sec:model}

\begin{figure}[!htb]
    \centering
    \includegraphics[width=\columnwidth]{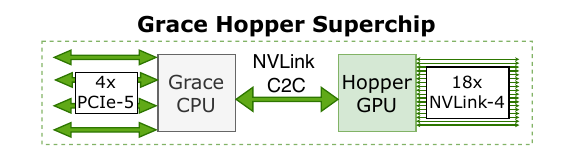}
    \caption{Grace Hopper Superchip.}
    \label{fig:model:ghsuperchip}
\end{figure}

To model the DGX~GH200 architecture in our network simulator, we first introduce the main components of the system. We begin by modeling a Grace Hopper Superchip, as illustrated in Figure~\ref{fig:model:ghsuperchip}. This Superchip consists of a Grace CPU connected to a generic intra-node network via a PCIe~5 interface with four lanes, and to a Hopper GPU through an NVLink-C2C interconnect operating at a maximum rate of \num{3600}~Gbps. The Hopper GPU connects to the Grace CPU through NVLink-C2C and to the NVLink network through eighteen NVLink~4 lanes, each operating at up to \num{200}~Gbps, providing a total bandwidth of \num{3600}~Gbps per Hopper GPU.

\begin{figure}[!htb]
    \centering
    \includegraphics[width=\columnwidth]{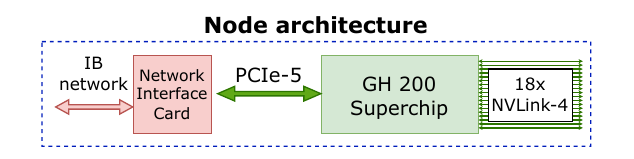}
    \caption{Node architecture.}
    \label{fig:model:nodeArch}
\end{figure}

This GH200 Superchip is then integrated into a general \textit{node} architecture, where the Superchip connects to the NVLink network through the eighteen NVLink~4 lanes. On the other side, the GH200 Superchip interfaces with a network interface card (NIC) via the PCIe connection, allowing the node to connect to an InfiniBand network. In our experiments, we focus exclusively on the NVLink network, neglecting the InfiniBand interconnect. Figure~\ref{fig:model:nodeArch} illustrates this node architecture.

\begin{figure}[!htb]
    \centering
    \includegraphics[width=0.75\columnwidth]{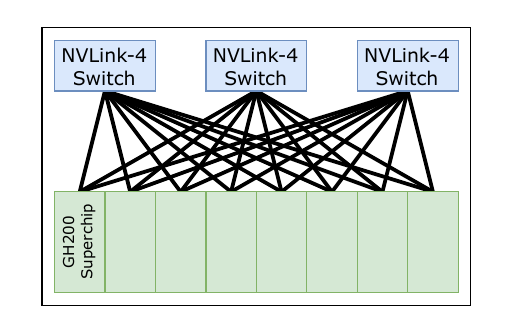}
    \caption{Compute tray architecture.}
    \label{fig:model:computeTray}
\end{figure}

Within the NVLink network, a DGX~GH200 system can be configured with up to \num{32} compute trays, each containing eight GH200 Superchips. Each NVLink~4 connection operates at \num{200}~Gbps, and each GH200 Superchip connects to three different NVLink~4 switches. Each Superchip establishes six links to each NVLink~4 switch, providing a total of \num{1200}~Gbps per Superchip. Consequently, each NVLink~4 Level~1 (L1) switch aggregates \num{48} NVLink~4 connections from Superchips, enabling high-bandwidth communication within each compute tray. Figure~\ref{fig:model:computeTray} illustrates the compute tray architecture.

\begin{figure}[!htb]
    \centering
    \includegraphics[width=\columnwidth]{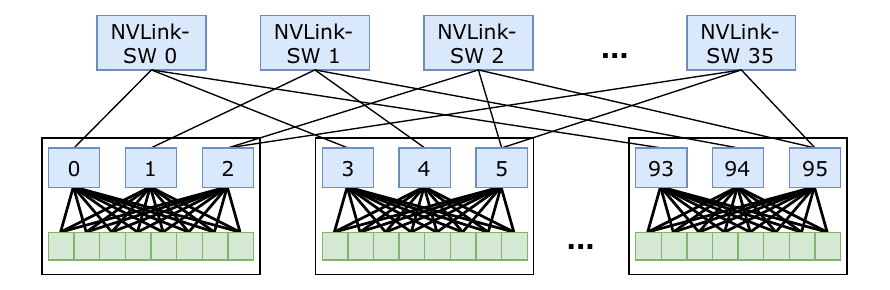}
    \caption{DGX~GH200 architecture.}
    \label{fig:model:dgxGH200Arch}
\end{figure}

Finally, a complete DGX~GH200 system with \num{256} GH200 Superchips comprises \num{32} compute trays, up to \num{96} L1 switches, and \num{36} Level~2 (L2) switches. Each L1 switch connects to twelve different L2 switches (one for every three L1 switches) using two links of \num{200}~Gbps each, achieving a total of \num{4800}~Gbps of bandwidth per L1 switch toward the L2 layer. Figure~\ref{fig:model:dgxGH200Arch} illustrates the overall DGX~GH200 architecture.

\section{Simulation results}
\label{sec:experiments}

In this section, we analyze four different configurations (\num{32}, \num{64}, \num{128}, and \num{256} GPUs), corresponding to the available configurations of the DGX GH200 system.

The bandwidth of the DGX GH200 system depends on the number of GPUs and NVLink Switches that compose the overall architecture. Each GH200 superchip provides a total bandwidth of \num{3600}~Gbps (\num{1200}~Gbps from each superchip to a first-level switch). Therefore, the bandwidth between the superchips and the first-level switches depends solely on the number of GPUs. In contrast, the bandwidth between the first- and second-level switches depends on the number of first-level switches. Since each first-level switch exposes \num{400}~Gbps to each second-level switch, the total bandwidth between these two switch levels is determined exclusively by the number of first-level switches. Table~\ref{tab:exp:dgxConfig} shows the possible configurations and bandwidth.

\begin{table}[!htb]
    \centering
    \caption{DGX~GH200 interconnect configuration and bandwidth.}
    \label{tab:exp:dgxConfig}
    \begin{tabular}{@{} l | c | c | c | c @{}} 
        \toprule
        \textbf{Number of GPUs} & \textbf{32} & \textbf{64} & \textbf{128} & \textbf{256} \\ 
        \midrule
        \textbf{L1 Switches} & \num{12} & \num{24} & \num{48} & \num{96} \\ 
        \textbf{Bandwidth GPU-L1 (Tbps)} & \num{115.2} & \num{230.4} & \num{460.8} & \num{921.6} \\
        \midrule
         \textbf{L2 Switches} & \num{36} & \num{36} & \num{36} & \num{36} \\
        \textbf{Bandwidth L1-L2 (Tbps)} & \num{57.6} & \num{115.2} & \num{230.4} & \num{460.8} \\ 
        \bottomrule
    \end{tabular}
\end{table}

\begin{figure}[!htb]
    \centering
    \begin{subfigure}{\columnwidth}
        \centering
        \frame{\includegraphics[width=1\textwidth]{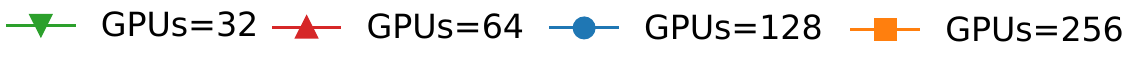}}
    \end{subfigure}
    \\
    \begin{subfigure}{\columnwidth}
        \centering
        \includegraphics[width=1\textwidth]{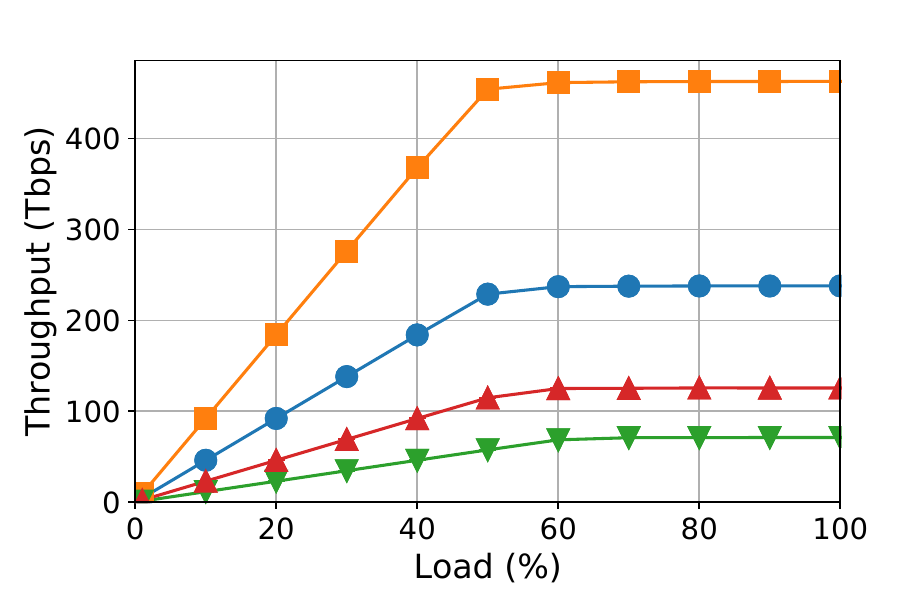}
    \end{subfigure}
    \caption{DGX GH200 cluster performance as a function of traffic load (\%).}
    \label{fig:exp:dgx}
\end{figure}

Figure~\ref{fig:exp:dgx} presents the results for these four configurations on a DGX GH200 cluster. The experiments were configured using a random all-to-all traffic pattern, where each GH200 superchip generates traffic according to its load. The traffic generated by each GH200 superchip varies from \num{0}\% (\num{0} Gbps) to \num{100}\% (\num{3600} Gbps). In the figure we can observe how the system achieved a maximum throughput of \num{450}~Tbps while using the whole system, being a good alternative to AI workloads, where the traffic is intensive. The four different allowed configurations saturate over the same traffic load, near to \num{50}~\% of the possible load. It is due to the network topology, an slimmed fat-tree, which achieves its maximum throughput when the communication is produced into individual chassis of \num{8}~GPUs.

\bibliographystyle{IEEEtran}
\bibliography{bib}
\end{document}